\documentclass[useAMS,usenatbib,usegraphicx]{mn2e}

\def\aap{{A\&A}}		
\def\apj{{ApJ}}			
\def\apjl{{ApJ}}		
\def\apjs{{ApJS}}		
\def\aj{{AJ}}			
\def\aaps{{A\&AS}}		
\def\pasp{{PASP}}		
\def\apss{{Ap\&SS}}		
\def\iaucirc{{IAU~Circ.}}       
\def\mnras{{MNRAS}}		
\def\pasj{{PASJ}}		

\newcommand{\erg}{{\rm\thinspace erg}}
\newcommand{\s}{{\rm\thinspace s}}
\newcommand{\ergps}{\mbox{$\erg\s^{-1}$}}

\title[Orbital Period Determinations for Four SMC BeXRBs]{Orbital Period Determinations for Four SMC Be/X-ray Binaries}
\author[M.P.E. Schurch et al.]{M.P.E. Schurch$^{1}$\thanks{E-mail: schurch@ast.uct.ac.za}, M.J. Coe$^{2}$,  V.A. McBride$^{2}$, L.J. Townsend$^{2}$, A. Udalski$^{3}$, 
\newauthor F. Haberl$^{4}$ and R.H.D. Corbet$^{5}$\\
$^{1}$Astrophysics, Cosmology and Gravity Centre (ACGC), Astronomy Department, University of Cape Town,\\
\,\,Rondebosch, Private Bag X1, 7701, South Africa.\\
$^{2}$School of Physics and Astronomy, Southampton University, Highfield, Southampton, SO17 1BJ, UK.\\
$^{3}$Warsaw University Observatory, Aleje Ujazdowskie 4, 00-478 Warsaw,Poland.\\
$^{4}$Max-Planck-Institut f\"ur extraterrestrische Physik, Giessenbachstra\ss{}e, 85748, Germany.\\
$^{5}$University of Maryland Baltimore County, X-ray Astrophysics Laboratory, Mail Code 662, NASA Goddard Space Flight Center,\\
\,\,Greenbelt, MD 20771, USA.}
\begin{document}

\date{26 October 2010, accepted for publication in MNRAS}

\pagerange{\pageref{firstpage}--\pageref{lastpage}} \pubyear{2010}

\maketitle

\label{firstpage}

\begin{abstract}
We present an optical and X-ray study of four Be/X-ray binaries located in the Small Magellanic Cloud (SMC).  OGLE I-band data of up to 11 years of semi-continuous monitoring has been analysed for SMC X-2, SXP172 and SXP202B, providing both a measurement of the orbital period ($P_{orb} = 18.62, 68.90,$ and $229.9$ days for the pulsars respectively) and a detailed optical orbital profile for each pulsar.  For SXP172 this has allowed a direct comparison of the optical and X-ray emission seen through regular {\em RXTE} monitoring, revealing that the X-ray outbursts precede the optical by around 7 days.  Recent X-ray studies by {\em XMM-Newton} have identified a new source in the vicinity of SXP15.3 raising doubt on the identification of the optical counterpart to this X-ray pulsar.  Here we present a discussion of the observations that led to the proposal of the original counterpart and a detailed optical analysis of the counterpart to the new X-ray source, identifying a 21.7\,d periodicity in the OGLE I-band data.  The optical characteristics of this star are consistent with that of a SMC Be/X-ray binary.  However, this star was rejected as the counterpart to SXP15.3 in previous studies due to the lack of H$\alpha$ emission.
\end{abstract}


\section{Introduction}
The Small Magellanic Cloud has been a hot bed for the discovery of High Mass X-ray Binaries ever since the initial discovery of SMC X-1 \citep{Leong71}. There are 56 known X-ray pulsars in the SMC \citep{coe05, Shtykovskiy05, haberl08b} and more than 20 candidates discussed in the literature.  These systems comprise a compact object, typically a neutron star, accreting mass from an early type companion star.  In all but one SMC system (SMC X-1 accretes from a supergiant) this companion is a main sequence Be  star, such systems are known as Be/X-ray binaries.  Typically these systems are found to have long period, high eccentricity orbits, that are directly related to the spin period of the neutron star through the Corbet diagram \citep{Corbet84}. The nature of such orbits leads to two types of outburst behaviour, Type-I outbursts occur during periastron when the neutron star interacts with the Be star's circumstellar disk, hence these outbursts are periodic and last a few days depending on the orbital eccentricity \citep{okazaki01}.  Type-II outbursts are far longer in duration and brighter, they can appear at any orbital phase and are associated with massive disk growth allowing continued mass transfer to the neutron star.  Studies of these systems and outbursts have been through their emission in X-rays and optical wavelengths. The high spatial capabilities of satellites like {\em XMM-Newton}, {\em Chandra} and {\em SWIFT}, allowed the counterparts to be identified as Be stars.  Recent optical studies by \citet{mcbride08} and \citet{antoniou09a} have produced optical classifications for the majority of the identified optical counterparts to both the pulsating and non-pulsating X-ray binaries in the SMC.  These studies show that the spectral class distribution is entirely consistent with that of the Milky Way Be/X-ray binaries.

\begin{figure}
 \includegraphics[width=83mm, angle=0]{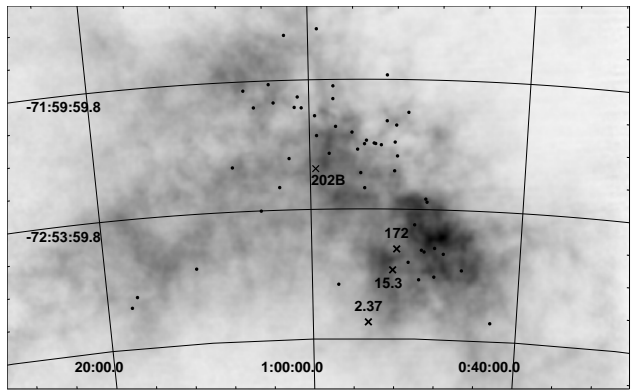}
 \centering
  \caption{HI map of the SMC \citep{Stanimirovic99} with the pulsars discussed in the paper marked with crosses and their pulse period.  Marked by black dots are other known SMC Be/X-ray binaries.\label{fig:source_positions}}
\end{figure}

\section{Data}
\subsection{Optical Monitoring}
Variations in the size of the circumstellar disk are primarily studied through H$\alpha$ spectroscopy.  However, the circumstellar disk also adds to the general continuum emission, this contribution is most noticeable at infrared wavelengths due to the relatively cool disk temperatures.  Studies of circumstellar disk temperatures have found that the disk surface temperature profile is extremely complex and highly dependent on density.  The range of temperatures is found to be $T_{disk} \approx (0.6-1.6)\times10^{4}$\,K \citep{Sigut09}.  As a result, the OGLE I band data is ideal for studying the changes that take place in Be stars' circumstellar disks.  The OGLE data primarily comprise two data sets, OGLE-II \citep{udalski97} and OGLE-III \citep{Udalski08a}, together covering the range MJD\,$50600-54665$. The OGLE data are sampled daily, but, due to the Earths orbit round the sun and the location of the OGLE telescope at the Las Campanas astronomical site in Chile, the SMC is only visible for approximately two thirds of the year.  Hence the combined OGLE data (which will be referred to as the OGLE light curve) has large periodic gaps. In addition MACHO light curves \citep{alcock99} have been used to supplement this study when optical data are required prior to the OGLE light curves.  The MACHO observations provided continuous monitoring with approximately daily sampling during the period MJD\,$49150-51550$.

\begin{table*}
 \centering  
 \begin{minipage}{155mm}
  \caption{Optical counterpart identifications proposed. The orbital period ($P_{orb}$) and ephemeris ($T_{0}$) are the values proposed in this paper.\label{ta:chap3_sources}}
  \begin{tabular}{@{}llllllll@{}}
  \hline
  \noalign{\smallskip}
   Name & R.A. & Dec. & MACHO$^{1}$ & OGLE-II$^{2}$ & OGLE-III$^{3}$ & $P_{orb}$ (days) & $T_{0}$ (MJD)$^{4}$ \\
  \hline
  \noalign{\smallskip}
   SMC X-2 & 00:54:34 & -73:41:03 & - & - & SMC107.5 25 & $18.62\pm0.02$ & $53170.7\pm0.2$\\
   SXP15.3-A$^{5}$ & 00:52:14 &	-73:19:19 & 212.16075.13 & SMC-SC6-99923  & SMC100.1 48026 & $75.1\pm0.5$ & $503761\pm1$ \\
   SXP15.3-B & 00:52:15 & -73:19:15 & 212.16075.79 & SMC-SC6-99991   & SMC100.1 48068 & $21.68\pm0.01$ & $50603.9\pm0.2$ \\
   SXP172 &  00:51:52 & -73:10:35 & 212.16077.13 & SMC-SC6-22749   & SMC100.2 44100 & $68.90\pm0.17$ & $50581.2\pm0.7$ \\
   SXP202B & 00:59:29 & -72:37:03 & 207.16541.15 & SMC-SC8-139407  & SMC105.3 29894 & $229.9\pm0.9$ & $52672\pm2$ \\
  \hline
  \noalign{\smallskip}
  \end{tabular}\\ 
  $^{1}$\citet{alcock99}, $^{2}$\citet{udalski97}, $^{3}$ \citet{Udalski08a}, $^{4}$ Ephemerides calculated for time of maximum brightness, $^{5}$ [MA93]552.
\end{minipage}
\end{table*}

\subsection{X-ray Monitoring}
The OGLE data are extremely well-matched to the X-ray monitoring of the SMC with the {\it Rossi X-ray Timing Explorer} space observatory \citep[{\em RXTE},][]{Jahoda96}. The Proportional Counter Array (PCA) instrument on board {\em RXTE} has been observing the SMC since 1997, full details of these observations and analysis are found in \citet{laycock05,galache08}.  Positional identification of SMC HMXBs is impossible with {\em RXTE} due to its 2$^\circ$ full width at zero intensity (FWZI) field-of-view (FoV).  Identification is performed solely through pulse timing analysis and identification with related activity in optical light curves.  Simultaneous evaluation of the {\em RXTE} and OGLE light curves is extremely powerful in identifying the true orbital period and outburst cycles for Be/X-ray systems, as has been shown for pulsars like SXP18.3 \citep{schurch09} and SXP46.6 \citep{mcgowan07}.

In this paper we present a thorough study of the X-ray and optical light curves of a number of SMC Be/X-ray binaries. Table \ref{ta:chap3_sources} lists the sources studied and respective optical identifications. Their positions within the SMC are also marked in Figure \ref{fig:source_positions}.

\begin{figure} \hspace{1.0em}
 \includegraphics[width=60mm, angle=90]{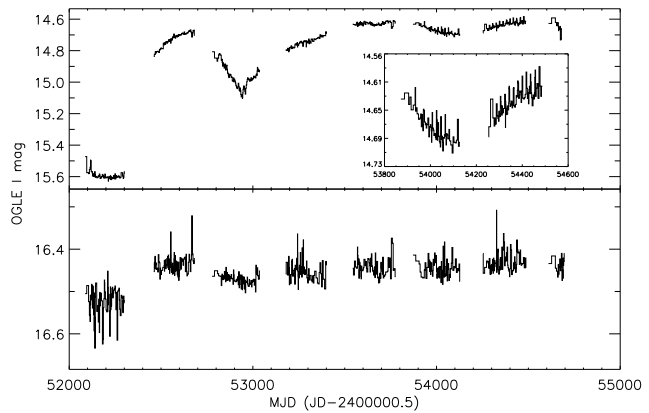}
  \caption{OGLE-III light curves for both the north (top) and south (bottom) candidates to SMC X-2.  The insert in the top panel shows an expanded view of the 6th and 7th year of OGLE-III data.\label{fig:SXP2.37_o_ns}}
\end{figure}

\section{SMC X-2 = SXP2.37}\label{sec:SXP2.37}
SMC X-2 also known by its pulse period identification SXP2.37 \citep[SMC X-ray Pulsar,][]{coe05}, was discovered in the SMC by SAS 3 observations in 1977 \citep{li77}.  Early X-ray observations showed the source to be highly variable \citep{li77a}.  Soon after the X-ray source was found, \citet{murdin79} revealed that the suspected optical counterpart was in fact a close north-south double with a separation of $\sim2.5$\arcsec; they classified the two stars as O (northern) and Be (southern). It has been impossible to identify clearly which of these two stars is the true counterpart due to the large X-ray error circle.  2.37\,s pulsations were first discovered during a giant outburst from January to May 2000 \citep{Corbet01}. 

\subsection{OGLE}
The two possible counterparts are not present in either the MACHO or OGLE-II catalogues.  \citet{Schmidtke06} presented a thorough analysis of the first 5 years of OGLE-III data for both stars.  They found no significant periodicities in either star and proposed the brighter more northern star to be the true counterpart due to the large variations in its light curve. \citet{mcbride08} classified the northern star as O9.5 III-V.

\begin{figure} \hspace{1.0em}
 \includegraphics[width=60mm, angle=90]{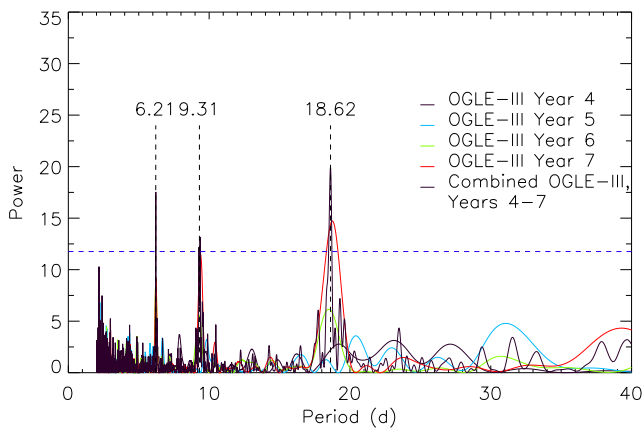}
  \caption{Periodogram of the OGLE-III data (4 years) for the northern counterpart to SMC X-2.  The coloured lines represent the individual years.  The black line is the  analysis of the entire 4 years of data.  The detected periodicity at 18.62\,d and its harmonics are marked.  The horizontal blue line represents the 99\% significance level.\label{fig:SXP2.37_n_ls}}
\end{figure} 

\begin{figure} \hspace{1.0em}
 \includegraphics[width=60mm, angle=90]{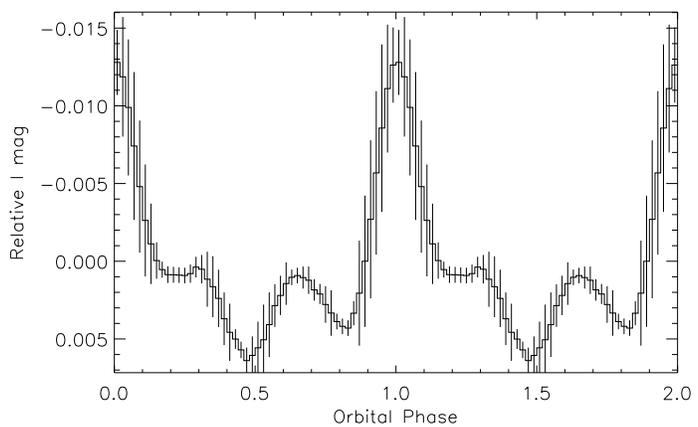}
  \caption{Optical orbital profile for the northern star counterpart to SMC X-2. Detrended OGLE-III data for years 4 to 7 were folded on 18.62\,d.\label{fig:SXP2.37_flc}}
\end{figure}

Two further years of OGLE-III data for both stars are now available providing a total coverage of 7 years. The total OGLE-III light curve for both the northern and southern objects is presented in Figure \ref{fig:SXP2.37_o_ns}.  It is clear to see that the northern object is highly variable varying by up to 1\,mag.  The insert in Figure \ref{fig:SXP2.37_o_ns} clearly shows that there is a periodicity in the 7th year of data. By comparison, the photometry of the southern star is fairly flat, with the variation in magnitude between the first and second year of data highly suggestive of contamination from the northern star. 

In order to search for periodicities in this data, it is essential first to detrend the data. However, it was decided to exclude the first three years of data for the northern star because the changes were too rapid to remove satisfactorily. The detrending was performed by fitting a straight line to each individual year of data. The two remaining light curves were then searched, both as a whole and in individual years, using Lomb-Scargle analysis \citep{Lomb76}. Figure \ref{fig:SXP2.37_n_ls} shows the analysis of the northern star.  The 99\% significance level for the data was calculated based on the false alarm probabiltiy (FAP) of finding a peak above a certain power in a pure noise light curve \citep{horne86}, resulting in the significance being given by 1-FAP.  The related error on the detected peak is given by Equation~14 in \citet{horne86}.  It is clear that in the 7th year of data a strong periodicity is present above the 99\% significance level.  The 2nd and 3rd harmonics are also identifiable, at a slightly lower level. In the analysis of the entire 4 years worth of data, the primary peak and both harmonics have increased in strength.  The periodicity is at $18.62\pm0.02$\,d with harmonics at 9.31 and 6.21\,d.

The same analysis was performed on the light curve of the southern star.  There is no periodicity seen in any of the individual 7 years of data above the 99\% significance level, and none when the light curve is taken as a whole.  This indicates that the observed periodicity in the northern star is real and not an artefact due to the close proximity of the stars, or of the data extraction process.  \citet{Schmidtke09} propose that the 18\,d period is an artefact of the timing analysis caused by the beating of the 6 and 9 day periodicities.  The visual presence of the 18.62\,d periodicity in the 7th year of data would suggest that beating is not the cause of the power and that the variation is infact due to the orbital motion, in addition the folded light curve (discussed later) is typical of the optical emission from a Be/X-ray binary system. 

Combining this period with the pulse period of 2.37\,s places SMC X-2 on the edge of the distribution of Be/X-ray binaries in the Corbet diagram \citep{Corbet09cp}. If one of the 6 or 9 day periodicites suggested by \citet{Schmidtke09} were the true binary period then the source would be fall within the Roche lobe overflow region (RLOF).  RLOF systems are continious accretors, hence they would be detectable in the X-ray at all times.  Although there is little X-ray information on this source (see Section \ref{sec:sxp2.37_rxte}) there have been sufficient observations of the region with {\em RXTE} to rule out the possibility that the source is a RLOF system.  This supports the proposal that 18.6\,d is the binary period \citep{Schurch08a} and that SXP2.37 is a Be/X-ray binary with the northern star as the optical counterpart \citep{Schmidtke06}.  Figure \ref{fig:SXP2.37_flc} shows the orbital profile for SMC X-2, constructed from the portion of optical light curve used in the period search.  In this paper the orbital profiles are constructed using the phase-independent folding procedure \citep{deJager88, galache08}. The data are folded a total of 5 times using 10 bins, each individual fold has a small starting point offset.  The combined folded profile is the average of the individual folds, producing a higher resolution profile. Although only every 5th bin is independent this has the effect of smoothing out spurious data points across several bins whilst allowing true variations to be enhanced.  The orbital profile reveals a very steep rise and fall in optical brightness.  This type of emission is a typical signature of a neutron star passing through periastron, the measured period would be that of the orbital period. There is also an additional emission peak at phase 0.65. This secondary peak could be due to increased accretion as extra material is picked up as the neutron star turns around and begins its next approach.  The ephemeris derived for the time of maximum brightness is MJD $(53170.7\pm0.2) + n(18.62\pm0.02)$\,d.

\subsection{{\em RXTE}}\label{sec:sxp2.37_rxte}
SMC X-2 is located in the south western corner of the SMC towards the outer edges (see Figure \ref{fig:source_positions}).  Due to this position and the observing strategy of {\em RXTE} \citep[see][]{galache08} it has rarely fallen within {\em RXTE}'s FoV, as a result little is known about its X-ray activity.  The only X-ray detection to overlap with the OGLE-III coverage was recorded on MJD 52228.  There is no noticeable correlation between this X-ray detection and any optical feature of either star that would aid in identifying the correct counterpart. Full analysis of the X-ray light curve is presented in \citet{galache08}.  There is no supporting orbital period from the X-ray data. 

\section{SXP15.3}\label{sec:SXP15.3}
\begin{figure}
 \includegraphics[width=83mm, angle=0]{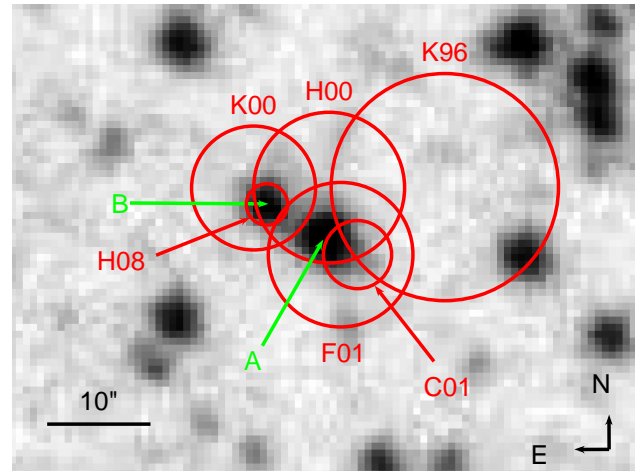}
 \centering
  \caption{ESO MAMA U-band image of SXP15.3 centred on the position of [MA93]522 (R.A.= 00:52:14, Dec.=-73:19:18).  The red circles refer to the X-ray positions and errors given in Table \ref{ta:SXP15.3_error}, the lables refer to the short name given in Table \ref{ta:SXP15.3_error}.  Labelled in green are the two identified counterparts, A ([MA93]522) is the current counterpart to SXP15.3, and B is the counterpart to the {\em XMM-Newton} source.\label{fig:SXP15.3_error}}
\end{figure}

As part of an analysis of recent {\em XMM-Newton} data, \citet{haberl08b} reported a source at the position R.A.=00:52:15.3, Dec.=-73:19:14.6 (J2000) with a total error radius of 1.93\arcsec and identified it as RX J0052.1-7319 (otherwise known as SXP15.3).  The position of this source has now raised issues over the correct optical counterpart for SXP15.3. Here the refinements made to the X-ray positions since the discovery of the source in 1996 are described, including the history behind the identification of an optical counterpart.  Table \ref{ta:SXP15.3_error} contains the X-ray positions and errors, these are also plotted over the ESO MAMA U-band image in Figure \ref{fig:SXP15.3_error}.  

\begin{table*}
 \centering
 \begin{minipage}{120mm}
  \caption{X-ray positions for SXP15.3 \label{ta:SXP15.3_error}}
  \begin{tabular}{@{}p{3.7cm}lrrrr@{}}
  \hline
  \noalign{\smallskip}
   Reference & Detector & R.A. & Dec. & r$_{90}$ (\arcsec) & Short Name$^{1}$\\
  \hline
  \noalign{\smallskip}
  \citet{kahabka96} & {\em PSPC} & 00:52:11.30 & -73:19:13.0 & 11.0 & K96\\
  \citet{kahabka00} & {\em HRI}  & 00:52:15.60 & -73:19:13.0 & 6.0  & K00\\
  \citet{haberl00} & {\em PSPC} \& {\em HRI} & 00:52:13.90 & -73:19:13.0 & 7.3 & H00\\
  \citet{covino01} & {\em HRI} & 00:52:13.27 & -73:19:19.5 & 3.3 & C01\\
  \citet{finger01} & {\em HRI} & 00:52:13.65 & -73:19:19.5 & 7.1 & F01\\
  \citet{haberl08b} & {\em XMM-Newton} & 00:52:15.30 & -73:19:14.6 & 1.93 & H08\\
   \hline
   \noalign{\smallskip}
  \end{tabular}\\
  $^{1}$This name is used in Figure \ref{fig:SXP15.3_error}.
\end{minipage}
\end{table*}  

\citet{lamb99} reported detecting pulsations at 15.3\,s from the X-ray source RX J0052.1-7319 in both {\em ROSAT} {\em HRI} and {\em BATSE} observations taken in 1996. The likely optical counterpart was then subsequently identified by \citet{israel99} as a Be star with R=14.54 that was on the edge of the current 11\arcsec {\em ROSAT} {\em PSPC} X-ray error circle \citep{kahabka96}, star A.  This identified optical counterpart is found in the optical catalogues of \citet{Meyssonnier93} as [MA93]552.  However, the authors also note the presence of an R=16.1 star just within the error circle, the identification of this star is unclear as no coordinates were reported. \citet{kahabka00} refined the {\em ROSAT} error through careful analysis of several {\em ROSAT} {\em HRI} observations. This new position is incompatible with their previously reported position, as can be seen in Figure \ref{fig:SXP15.3_error}.  From this new error circle the authors propose two potential optical candidates.  These two candidates are [MA93]552 and a fainter star to its northeast (stars A and B respectively in Figure \ref{fig:SXP15.3_error}).  They analysed early OGLE-II data revealing that both stars were variables.  The authors also reported that no H$\alpha$ emission has been detected from the fainter northern star (B).  In light of this, they supported the identification of the brighter Be type star [MA93]522 (A) as the counterpart.  At the same time as these data were published, \citet{haberl00} analysed both {\em ROSAT} {\em PSPC} and{\em HRI} data.  Their position is consistent with star A and so they supported the identification of [MA93]522 as the counterpart.  However, it is clear from Figure \ref{fig:SXP15.3_error} that both candidates are within this error circle. \citet{covino01} then presented a detailed analysis of the {\em ROSAT} {\em HRI} data using a sliding cell and wavelet transform based algorithm.  After a subsequent boresight correction, their final position had an error radius of 3.3\arcsec, providing the most precise position.  The error circle now clearly selects [MA93]552 as the only possible optical counterpart, with star B clearly outside the error circle.  They performed spectroscopic and photometric analysis of [MA93]552, finding a classification of either O9.5IIIe or B0Ve depending on the extent of reddening that can be attributed to the circumstellar disk.  They also noted that star B shows no sign of emission lines at H$\alpha$ or H$\beta$, and has a spectral class that is later than [MA93]552.  \citet{finger01} presented the timing analysis of the original data taken in 1996 when the source was at its brightest. They supported the position and identification of the counterpart made by \citet{covino01} and suggested that the earlier positions quoted by \citet{kahabka00} were less accurate due to the low count rate of the data they analysed.

The {\em XMM-Newton} detection is the first observation of this region by a telescope with a spatial accuracy that is good enough to distinguish clearly between the two possible candidate stars.  The combined 1$\sigma$ error and systematic error produce an error radius of 1.93\arcsec (see Figure \ref{fig:SXP15.3_error}) which clearly selects only one possible counterpart star.  This star is not the previously identified optical counterpart to SXP15.3 ([MA93]522), but is infact associated with the discarded candidate star B.  The {\em XMM-Newton} source was detected with a luminosity of $\sim L_{0.2-10}=4.0\times10^{34}$\ergps, this level of X-ray activity is typical of Be/X-ray binaries in a quiescent state.  The $\sim$300 counts were insufficient to detect any X-ray periodicities and hence the association with SXP15.3 is not certain.  An X-ray upper flux limit at the position of [MA93]522 from this observation is a factor two lower at $\sim L_{0.2-10}=2.0\times10^{34}$\ergps.  Two possibilities remain:

\begin{itemize}
\item That the detected {\em XMM-Newton} source is the same source that {\em ROSAT} detected 15.3\,s pulsations from, and that the previous attempts to reduce the systematic errors in the {\em ROSAT} observations were incorrect.  The counterpart is the northern star (B), with a spectral classification later than B0V.
\item That there are two distinct X-ray sources associated with two separate optical counterparts.  The 15.3\,s pulsations are associated with the Be star [MA93]552.
\end{itemize}

\begin{figure} \hspace{1.0em}
 \includegraphics[width=60mm, angle=90]{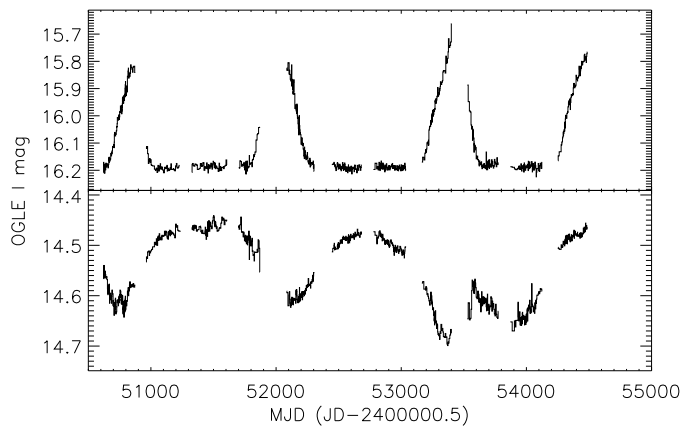}
  \caption{OGLE light curves of candidate counterparts to RX J0052.1-7319. [MA93]552 (bottom) identified by {\em ROSAT}, and the northern candidate (star B, top) identified by {\em XMM-Newton}.\label{fig:SXP15.3_both}}
\end{figure}

\subsection{OGLE}
\citet{EdgePhD} and \citet{coe05} have extensively analysed a total of 11 years of MACHO and OGLE data for [MA93]552. They identified a possible orbital period of 75\,d.  The subsequent 4 years of OGLE data continue to show the large optical variability of this source (Figure \ref{fig:SXP15.3_both}).  It was decided not to re-analyse this data due to the difficulty in removing the extremely high level of variability.  Here we present the optical analysis of the merged OGLE-II and III data (identifications are given in Table \ref{ta:chap3_sources}) for the northern counterpart star B that is firmly associated with the {\em XMM-Newton} source, the MACHO data was not included in the analysis due to the high noise levels caused by the larger PSF and contamination from [MA93]552.

Figure \ref{fig:SXP15.3_both} shows the extreme optical variations in the OGLE light curve.  The $\sim$0.5\,mag outbursts are semi-periodic, recurring on timescales of $1000-1400$\,d and lasting for up to 600\,d.  These timescales are far too large and too variable to be attributed to the orbital period of the system.  Similar outbursts are seen in other sources such as SXP18.3 \citep{schurch09,haberl08b} and are a result of changes in the structure and size of the circumstellar disk. Detrending the large outbursts was performed because small blips are evident that have been seen in other SMC Be/X-ray binaries \citep[for example, SXP46.6][]{mcgowan07} and are often characteristic of an orbital signature.  Due to the extremely linear rise and fall of the outbursts, the light curve was fully detrended using only linear fits.  The fifth year of data was removed from the final detrended light curve, due to the poor linear fit, producing an artificial wiggle in the light curve.

\begin{figure} \hspace{1.0em}
 \includegraphics[width=60mm, angle=90]{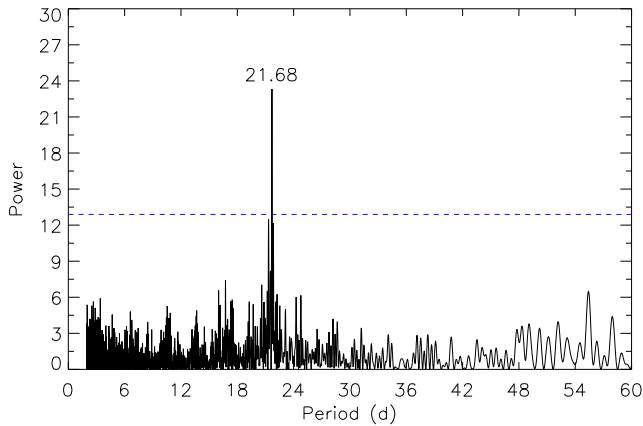}
  \caption{Periodogram of the detrended OGLE light curve of the potential northern counterpart star B to the source RX J0052.1-7319.  The horizontal blue line represents the 99\% significance level.\label{fig:SXP15.3_ls}}
\end{figure}

 \begin{figure} \hspace{1.0em}
 \includegraphics[width=60mm, angle=90]{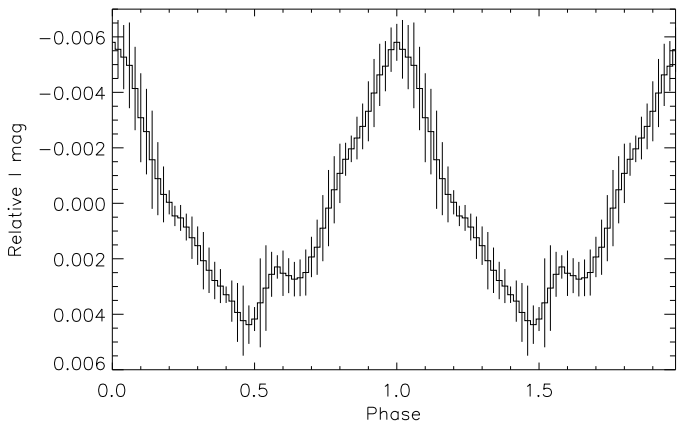}
  \caption{Optical orbital profile for SXP15.3 counterpart B.  The detrended OGLE-III light curve was folded on 21.68\,d.\label{fig:SXP15.3_flc}}
\end{figure}

The periodogram shown in Figure \ref{fig:SXP15.3_ls} clearly shows a significant peak well above the 99\% significance level in the data at $21.68\pm0.01$\,d.  The detrended optical light curve was folded at this period (Figure \ref{fig:SXP15.3_flc}) revealing a very symmetrical orbital profile with a steady rise and fall in brightness. If it is assumed that the correct scenario is that this star is the real counterpart to the 15.3\,s pulsating X-ray source, then the period found in the OGLE light curve would be entirely consistent with the expected orbital period from the Corbet diagram.  The system would occupy a position at the lower end of the distribution of Be/X-ray binaries on the Corbet diagram.  However, the previously identified counterpart to the southwest also exhibits an optical period that is consistent with the expected range of orbital periods.  An orbital period of 75\,d would place the system in the middle of the distribution of Be/X-ray binaries on the Corbet diagram.  Both stars seem to exhibit orbital periods and behavioural variations that are consistent with other counterparts to Be/X-ray binaries.  The optical magnitude ranges are also consistent with other counterparts in the SMC.  The only major difference between the two sources is the spectral classification.  The southern star ([MA93]552) has been well classified as a Be star as would be expected from a HMXB counterpart. From spectra taken on 19$^{th}$ and 20$^{th}$ January 1999 by the Danish 1.5\,m telescope at La Silla, \citet{covino01} noted that the northern star B shows no emission features and has a later type spectrum than [MA93]522.  However, this does not rule out the possibility it is a mid or even late B type star, as suggested by the spectrum in figure 2 of \citet{covino01}.  Through independent photometry taken on 20$^{th}$ January 1999 by the 1.0\,m SAAO telescope, \citet{finger01} also noted that the northern star shows no H$\alpha$ excess when compared to the surrounding stars, unlike the southern star.  These spectral and photometric characteristics are inconsistent with other HXMB counterparts. However, these observations were taken during the period MJD\,$51197-8$; Figure \ref{fig:SXP15.3_both} indicates that the star was in its base state at the time of these observations.  If this state is indicative of a B star without a circumstellar disk, then no H$\alpha$ emission would be expected.  Both optical spectroscopy during an optical outburst, to identify the circumstellar disk, and high resolution spectral classification of the northeastern star B would enable the true nature of this star to be identified and either placed with the typical Be star counterparts or separately as the first later type counterpart.

\subsection{{\em RXTE}}
\begin{figure} \hspace{1.0em}
 \includegraphics[width=60mm, angle=90]{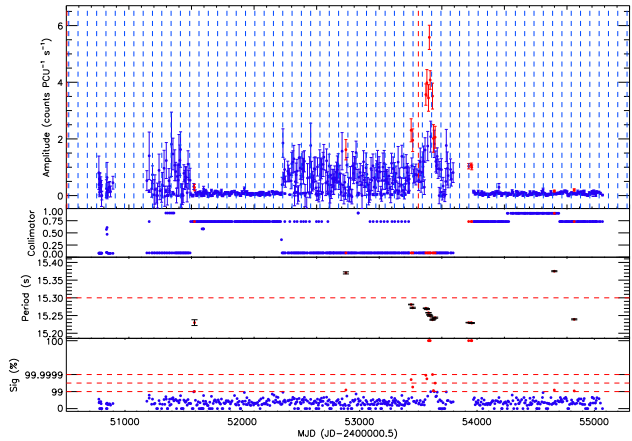}
  \caption{{\em RXTE} light curve of SXP15.3.  Panels are PCA pulsed flux light curve in the energy range 3-10 keV corrected for the collimator response (top panel, blue dashed lines indicate potential periastron passages based on the detected orbital period), PCA collimator
response (second from top), PCA detected pulse period for detections above 99\% significance (second from bottom, red dashed line indicates the standard period expected for the source) and PCA detection significance (bottom, red dashed lines show significance levels 99\%, 99.9\% and 99.9999\%).  The red points denote a detection significance above 99\% and the blue below.  \label{fig:SXP15.3_xlc}}
\end{figure}

The {\em RXTE} light curve for SXP15.3 is shown in Figure \ref{fig:SXP15.3_xlc}.  During the course of the monitoring program, detections of SXP15.3 have been rare.  However, for a period of about 100 days SXP15.3 did enter into a giant Type-II outburst phase (beginning on MJD\,53564).  This outburst was particularly interesting since it showed signs of being orbitally modulated.  Unsuccessful attempts to fit an orbital solution to the data were performed by \citet{galache08}; however, they note that variations in the detected period during the Type-II outburst suggest an orbital period of $\sim$28\,d.  Possible correlations were examined between the observed X-ray activity and the two possible optical light curves, in an attempt to distinguish the correct counterpart.  There are no obvious features in the optical light curves that seem to correspond with any of the X-ray activity.  \citet{EdgePhD} noted that an earlier X-ray detection by the {\em ROSAT} {\em HRI} occurred simultaneously with a large peak in the MACHO data for the southern star.  An independent Lomb-Scargle analysis of the {\em RXTE} light curve was performed.  No significant periodicities were found in the data.  \citet{galache08} suggested a period of $\sim$28\,d by fitting the pulse period changes observed during the large Type-II outburst with an orbital model.  This period is neither confirmed nor ruled out by our analysis of the {\em RXTE} light curve data.

\subsection{SXP15.3 Discussion}
Two potential optical candidates exist for SXP15.3.  The analysis of both optical light curves has been unable to clarify the identity of the true optical counterpart.  Through the {\em XMM-Newton} X-ray detection it is known that there is an X-ray source associated with the northern star B.  This northern star is suspected to be a binary system with a 21.7\,d orbital period. However, from the earlier spectra presented in \citet{covino01} this candidate star does not resemble a typical Be star counterpart. The association of this source with SXP15.3 is still unclear, and so it cannot be identified as a pulsating neutron star system.  The uncertainty over this source will remain until such a time that further X-ray observations are made during a period when SXP15.3 is in outburst. Such observations require simultaneous positional and timing capabilities so that the localisation and identification of the source can be made.  Optical observations of both potential X-ray binary systems are required, but in particular of the northern candidate during the next period of maximum optical flux, so that a precise optical classification can be made, allowing this source to be properly compared to the population of X-ray binary systems in the SMC.  Measurements of the H$\alpha$ emission line during this period will provide us with an indication as to whether the optical flux variations seen in the OGLE-III light curves are related to some sort of circumstellar disk activity.

\section{SXP172}\label{sec:SXP172}
SXP172 is a member of the group of pulsars in the south western region of the SMC.  Pulsed X-ray emission at 172.4\,s was discovered in {\em ASCA} data of the {\em ROSAT} source RX J0051.9-7311 \citep{Yokogawa00}.  The emission-line star [MA93]504 was suggested as the counterpart to these X-ray sources \citep{Cowley97}, with firm confirmation arising from the detection of 172.21\,s pulsed X-ray emission from the optical counterpart's location in an {\em XMM-Newton} observation \citep{haberl04}.  \citet{mcbride08} have classified the counterpart to be an O9.5-B0 V star, confirming its Be/X-ray binary classification.  The MACHO and OGLE-II data have been well studied by \citet{Schmidtke06a}.  They proposed an orbital period of 69.9\,d from the first two years of OGLE-II data.

\subsection{{\em RXTE}}
\begin{figure} \hspace{1.0em}
 \includegraphics[width=60mm, angle=90]{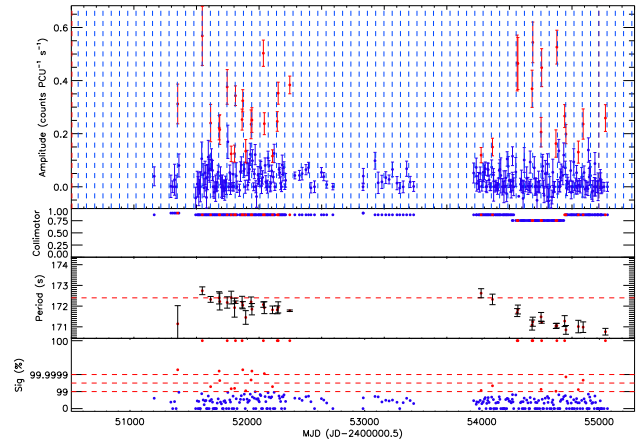}
  \caption{{\em RXTE} light curve of SXP172 where the source position was within 15\arcmin of the pointing position.  Panels are the same as Figure \ref{fig:SXP15.3_xlc}.\label{fig:SXP172>0.75_1}}
\end{figure}

{\em RXTE} has recently observed a series of seven outbursts from SXP172 over the course of $\sim$700\,days (Figure \ref{fig:SXP172>0.75_1}).  This is the second time in the history of SXP172 that a succession of X-ray outbursts has been seen. The outbursts occurring around MJD 52000 have previously been studied by \citet{laycock05} and \citet{galache08}.  The two analyses of these outbursts has suggested possible orbital periods at $\sim$67\,days and $\sim$70\,days respectively.  However, neither period determinations are conclusive \citep{galache08}.  Since this light curve was last analysed, two further years of data have been collected, and the second set of outbursts was observed.  Here we present the analysis of the longer light curve in order to determine the true orbital period of this Be/X-ray binary. 

\begin{figure} \hspace{1.0em}
 \includegraphics[width=60mm, angle=90]{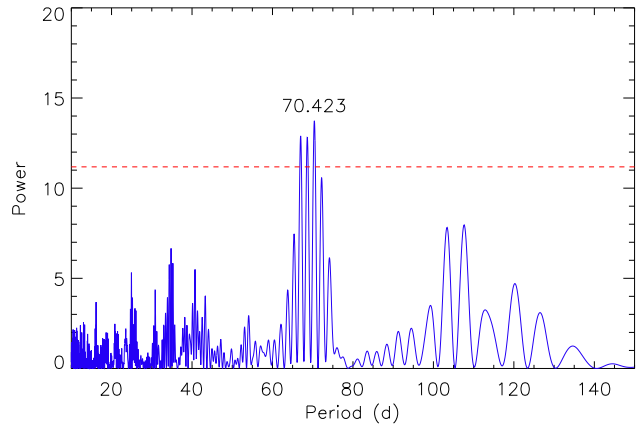}
  \caption{Lomb-Scargle analysis of {\em RXTE} light curve of SXP172. The horizontal dashed red line represents the 99\% significance level. \label{fig:SXP172>0.75_2}}
\end{figure}

As a direct result of the observing strategy of the {\em RXTE} monitoring program, there have been periods when SXP172 has fallen very close to the edge of the FoV of {\em RXTE}.  This reduced sensitivity causes all but the brightest outbursts to be missed.  As a result, including these periods of low collimator response will have a detrimental effect when searching for orbital periods using Lomb-Scargle analysis.  It was decided only to analyse data which were taken with SXP172 extremely close to the centre of the FoV (a collimator response of greater than 0.75), thus producing a data set with the maximum data quality.  The resultant X-ray light curve is shown in Figure \ref{fig:SXP172>0.75_1}.  Analysis of this light curve produces a number of significant peaks above the 99\% significance level as shown in Figure \ref{fig:SXP172>0.75_2}.  The highest of these peaks is at $70.42\pm0.15$\,d, which is consistent with the orbital period suggested by \citet{galache08} from the first period of outbursts, and with the optical period of \citet{Schmidtke06a}.  The other two significant peaks are at $68.68$\,d and $66.98$\,d.  It is believed that this complicated power structure is due to aliasing between the two slightly different periods found in the well separated outbursts when they are analysed separately.  When the outbursts are searched the maximum power for the first and second outbursts are found at $72.46$\,d and $68.49$\,d.  However, these periods are not significant on their own.  Until further X-ray outbursts are observed allowing the X-ray orbital period to be more precisely determined all three periods should be treated with equal significance.

\begin{figure} \hspace{1.0em}
 \includegraphics[width=60mm, angle=90]{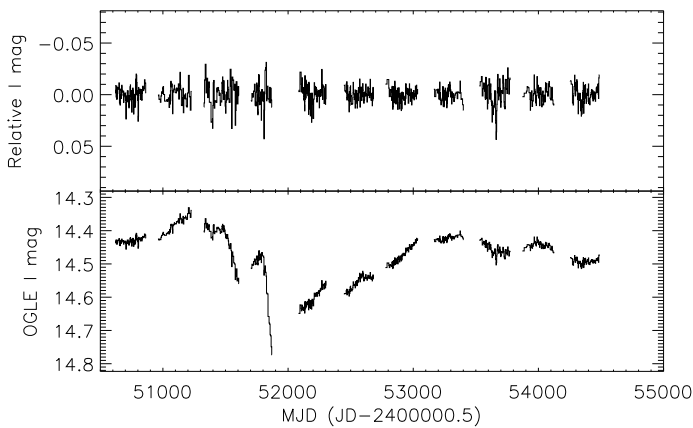}
  \caption{OGLE light curve of SXP172 (bottom) and detrended light curve top.\label{fig:SXP172_both}}
\end{figure}

\subsection{OGLE}
\citet{Schmidtke06a} report an orbital period of $69.9\pm0.6$\,d from analysing the first 2 years of OGLE-II data.  No periodicities were found in the corresponding MACHO or OGLE-III data.  They reported the period of 69.9\,d due to finding peaks in  both their Lomb-Scargle and phase dispersion
minimization (PDM) analyses at 23.4\,d and 34.9\,d, these form the second and third harmonics of a 69.9\,d period.  A re-analysis of the OGLE-II and OGLE-III data is presented in order to establish the orbital period.  The combined optical light curve (Figure \ref{fig:SXP172_both}) is extremely variable with sudden changes in observed brightness of up to 0.3\,mag.

\begin{figure} \hspace{1.0em}
 \includegraphics[width=60mm, angle=90]{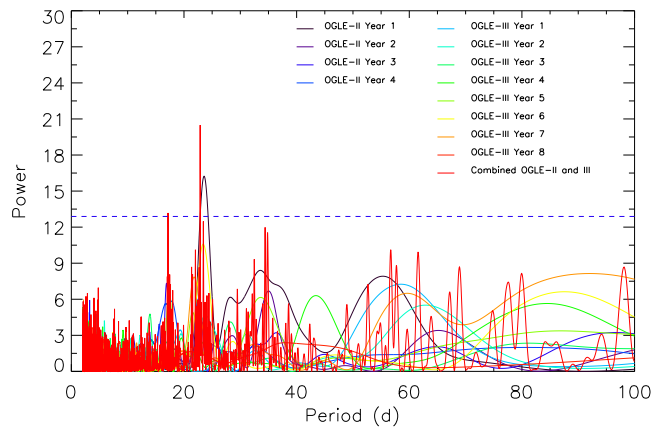}
  \caption{Lomb-Scargle analysis of detrended OGLE light curve of SXP172.  The horizontal blue line represents the 99\% significance level.\label{fig:SXP172_o_ls_det}}
\end{figure}

From the initial analysis of the light curve we recover the two peaks identified by \citet{Schmidtke06a} in the second year of OGLE-II data.  These peaks are below the 99\% significance level.  No further periodicities were seen in any of the subsequent years of data when either analysed separately or as a whole.  In light of this the light curve was detrended to remove the large magnitude changes in order to allow a fuller analysis of the smaller scale variations.  Detrending was performed using the method described in \S\ref{sec:SXP2.37} and \S\ref{sec:SXP15.3}.  This light curve (top of Figure \ref{fig:SXP172_both}), was then passed through exactly the same period analysis procedure.  The results shown in Figure \ref{fig:SXP172_o_ls_det} show that the previous peak at 34.9\,d is lost in the second year of the detrended data, but the peak at 23.4\,d gets slightly stronger and is now above the 99\% threshold.  When the entire light curve is searched as one (the red over-plotted line) it can be seen that there is a significant peak around 23\,d but it has shifted to a slightly shorter period of $22.9\pm0.1$\,d. The next peak also above the 99\% significance threshold occurs at 17.16\,d.  There is also a hint of the return of a peak at 34.4\,d slightly below the threshold.  These three values would form the second, third and fourth harmonics of a 68.9\,d period.  There is the suggestion of a peak around this value in the Lomb-Scargle periodogram.  However, it is extremely weak and would not be identifiable on its own.  A break down of each year shows no further signs of any periodicities.

Although the period found from the X-ray data differs slightly it is noted that one of the several potential peaks with a significance above 99\% in Figure \ref{fig:SXP172>0.75_2} sits at 68.7\,d.  In light of this close agreement between the derived periods from the independent analysis of both the X-ray and optical light curves, and the agreement of these values with the findings of \citet{Schmidtke06a}, it is proposed that the orbital period of the system is 68.9\,d.   

\subsection{SXP172 Discussion}
\begin{figure} \hspace{1.0em}
 \includegraphics[width=60mm,angle=90]{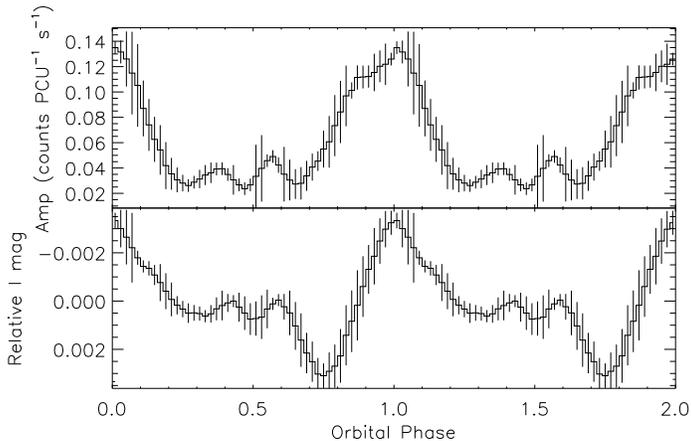}
  \caption{X-ray (top) and optical (bottom) orbital profiles for the counterpart for SXP172 folded on 68.9\,d.\label{fig:SXP172_flc_xte_opt}}
\end{figure}

Both the detrended optical and X-ray light curves for SXP172 were folded to compare the shape of the orbital profile.  The ephemeris found through the analysis of the optical data is MJD $(50581.2\pm0.7) + n(68.9\pm0.17)$\,d.  For a direct comparison, both the optical and X-ray light curves were folded using this ephemeris (Figure \ref{fig:SXP172_flc_xte_opt}).  It is clear that there is strong agreement between the two data sets.

The shape of the optical profile is unusual in that the optical emission seems to have one main peak and then a broad plateau lasting for approximately one third of the orbit.  This feature may be linked to the fact that the orbital period is detected through the presence of a strong second harmonic.  The main X-ray outburst is narrow and symmetric returning almost to a zero emission level.  This type of X-ray profile is typical of those seen in the other SMC Be/X-ray binaries \citep{galache08}.  The comparison with the optical reveals that the X-ray outburst begins 0.1 orbital cycles before the optical outburst, but they both peak at the same time.  This delay would suggest that the X-ray emission is the cause of the optical outburst, possibly through X-ray heating of the circumstellar disk.   Notable in both pulse profiles are two small peaks occurring around an orbital phase of 0.5.  Within errors these peaks are consistent with a straight line, but it is unusual that they appear in independent data sets with such good agreement.  Their presence could be linked to emission occurring as a result of the neutron stars passage through apastron.  Detailed observations throughout this orbital phase would allow the presence of these peaks to be confirmed. 

\section{SXP202B}\label{sec:SXP202B}
SXP202B, or XMMU J005929.0-723703, was recently discovered during a series of {\em XMM-Newton} observations of the SMC by \citet{haberl08b}.  {\it XMM-Newton's} positional accuracy allowed the identification of [MA93]1147 as the optical counterpart.  The counterpart was previously classified as B0-5(III)e in the 2dF survey of the SMC by \citet{evans04}. \citet{haberl08b} performed a thorough analysis of the MACHO and OGLE-II data revealing possible periodicities at 334\,d and 220\,d from broad peaks in the power spectra. These were attributed to quasi-periodic variations (QPVs) in the Be disk. 

\begin{figure} \hspace{1.0em}
 \includegraphics[width=60mm, angle=90]{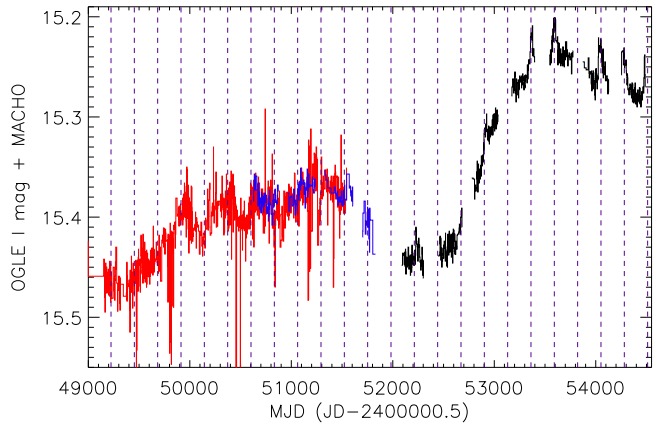}
  \caption{Optical light curve of SXP202B, MACHO (red) OGLE-II (blue) and OGLE-III (black). Vertical dotted lines show periods of predicted periastron passage.\label{fig:SXP202B_olc}}
\end{figure}

\begin{figure} \hspace{1.0em}
 \includegraphics[width=60mm, angle=90]{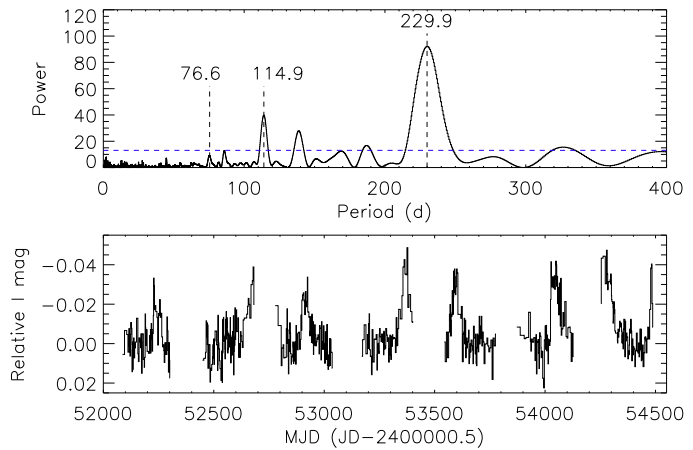}
  \caption{Detrended OGLE light curve of SXP202B (bottom).  The top panel shows the Lomb-Scargle analysis with the detected orbital period and harmonics marked.  The horizontal blue line represents the 99\% significance level. \label{fig:SXP202B_dolc_ls}}
\end{figure}

\subsection{OGLE}
The optical counterpart to SXP202B is an extremely variable source (Figure \ref{fig:SXP202B_olc}) showing dips and rises of the order 0.3\,mag.  On top of this underlying variation, there are 8 clear optical outbursts of $\sim$0.04\,mag in the OGLE-III light curve each lasting for around 100\,d.  From a visual check they appear to occur periodically.  The MACHO data hints at similar outbursts but, due to the data quality and underlying variations, they are hard to see.  It was decided to restrict the data analysis to the OGLE light curves. The data were detrended in order to remove the large variation.  Due to the significant length of the outbursts and their locations, the individual years were detrended by fitting a straight line to the data either side of the outbursts.  Figure \ref{fig:SXP202B_dolc_ls} shows the resultant OGLE light curve (bottom) and the Lomb-Scargle analysis (top). A strong periodicity at $229.9\pm0.9$\,d \citep{Schurch08} is revealed in the Lomb-Scargle periodogram.  Also identifiable are the 2nd and 3rd harmonics.  The two peaks at 140\,days and 85\,days are due to beating of both the fundamental and the 1st harmonic with the one year sampling. \citet{haberl08b} find a broad signal in the analysis of MACHO R-band data at 220\,d.  Combining the 229.9\,d period with the pulse period (202\,s) places the source directly in the centre of the distribution of Be/X-ray binaries on the Corbet diagram. 

\begin{figure} \hspace{1.0em}
 \includegraphics[width=60mm, angle=90]{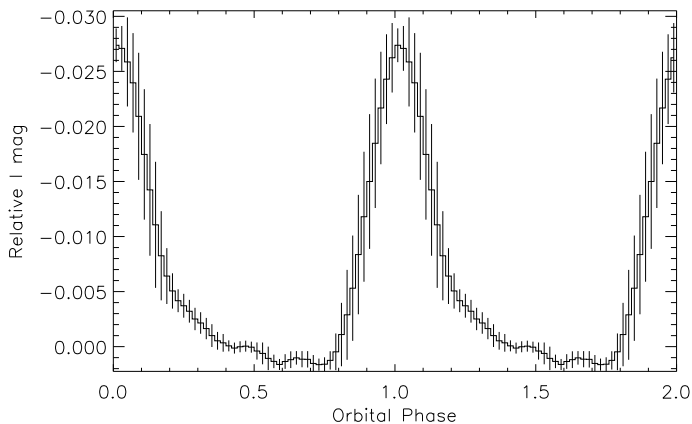}
  \caption{Optical orbital profile for SXP202B.  The detrended OGLE-III light curve was folded on 229.9\,d.\label{fig:SXP202B_oflc}}
\end{figure}

The orbital profile for SXP202B (Figure \ref{fig:SXP202B_oflc}) is typical of many of the SMC Be/X-ray binaries.  The profile is very symmetric with a steep rise and fall.  There is an extra tail in the decay of the outburst, which may suggest that the cause of the optical emission, possibly an over-density or hot spot in the circumstellar disk, slightly outlives the passage of the neutron star.  The ephemeris derived is MJD $(52672\pm2) + n(229.9\pm0.9)$\,d. 

\subsection{{\em RXTE}}
The location of SXP202B (approximately in the centre of the SMC) is extremely close to SXP202 originally discovered in {\em XMM-Newton} observations by \citet{majid04}.  The two sources were each identified and localised by {\em XMM-Newton} and are sufficiently distant from each other that there is no possibility that they are the same object.  Both sources routinely fall within the FoV of {\em RXTE}, as a result any extracted light curve \citep[see][]{galache08}) will be a combination of the two sources.  It was not possible to attribute any X-ray detections uniquely to either source based on their orbital periods.  In order to identify future X-ray outbursts to either pulsar we will require either a simultaneous optical outburst to be observed or to have follow-up X-ray observations with satellites combining both positional and timing capabilities, such as {\em SWIFT}, {\em XMM-Newton} or {\em Chandra}. 

\section{Conclusions}
The careful analysis of the optical and X-ray light curves for four SMC Be/X-ray binaries is now revealing a variety of behaviours. The MACHO and, in particular, OGLE I-band data have provided an extremely good optical history for these sources that is extremely well matched to the X-ray monitoring data of {\em RXTE}.  An 18.6\,d period has been identified in the northern counterpart to SMC X-2, strongly suggesting that it is the optical counterpart and that this system is a Be/X-ray binary.  The OGLE histories have revealed orbital periods for the known Be/X-ray binaries SXP172 and SXP202B at 68.90\,d and 229.9\,d respectively.  In addition we have identified a 21.7\,d orbital period in the optical counterpart to a new {\em XMM-Newton} source.  Due to the uncertainity in postion of both the {\em XMM-Newton} and {\em ROSAT} sources in this small patch of sky it is unclear whether there is one or two X-ray sources. The association of 15.3\,s X-ray pulsations detected by {\em ROSAT} with either source will require further X-ray and optical observations.  Future spectral classification of the new optical counterpart is necessary if its possible unusual spectral position within the growing population of SMC Be/X-ray binaries is to be evaluated properly.

\section*{Acknowledgments}
MPES is supported through the UCT/URC Postdoctoral Research Fellowship.  The OGLE project was partially supported by the Polish MNiSW grant N20303032/4275. LJT is supported by a Mayflower Scholarship from the University of Southampton.


\bsp

\label{lastpage}

\end{document}